\documentclass[nofootinbib, prd, twocolumn]{revtex4}
\usepackage{amsmath}
\usepackage{amsfonts}
\usepackage{amssymb}
\usepackage{dsfont}
\usepackage{mathrsfs}
\usepackage{bm,slashed}
\usepackage{graphicx}
\usepackage{color}
\usepackage{hyperref}
\usepackage{MnSymbol}
\hypersetup{
    colorlinks=true, 
    linktoc=page,    
    linkcolor=blue,  
    urlcolor=black
}

\DeclareFontFamily{U}{rsfs}{}         
\DeclareFontShape{U}{rsfs}{m}{n}{<5> rsfs5 <6><7> rsfs7          %
  <8><9><10><10.95><12><14.4><17.28><20.74><24.88> rsfs10}{}     %
\DeclareMathAlphabet{\mathfs}{U}{rsfs}{m}{n}                     %
\newcommand{\be}{\nopagebreak[3]\begin{equation}}
\newcommand{\ee}{\end{equation}}

\newcommand{\bee}{\nopagebreak[3]\begin{equation*}}
\newcommand{\eee}{\end{equation*}}

\newcommand{\ba}{\nopagebreak[3]\begin{eqnarray}}
\newcommand{\ea}{\end{eqnarray}}

\newcommand{\baa}{\nopagebreak[3]\begin{eqnarray*}}
\newcommand{\eaa}{\end{eqnarray*}}

\newcommand{\la}{\label}
\newcommand{\n}{\nonumber}

\newcommand{\R}{\mathbb{R}}
\newcommand{\Z}{\mathbb{Z}}

\def\pa{\partial}
\def\rd{\mathrm{d}}
\newcommand{\va}{\scriptscriptstyle}

\def\be{\begin{equation}}
\def\ee{\end{equation}}
\def\ba{\begin{eqnarray}}
\def\ea{\end{eqnarray}}

\def\label{\langle}

\def\su{\mathfrak{su}}

\def\ut#1{\rlap{\lower1ex\hbox{$\sim$}}#1{}}

\begin{document}

\title{
{ The loop gravity string}}
\author{Laurent Freidel}
\email{lfreidel@perimeterinstitute.ca}
\affiliation{Perimeter Institute for Theoretical Physics,
31 Caroline St. N, N2L 2Y5, Waterloo ON, Canada}

\author{Alejandro Perez}
\email{perez@cpt.univ-mrs.fr}
\affiliation{Aix Marseille Univ, Universit\'e de Toulon, CNRS, CPT, Marseille, France}

\author{Daniele Pranzetti}
\email{dpranzetti@sissa.it}
\affiliation{Scuola Internazionale Superiore di Studi Avanzati (SISSA), via Bonomea 265, 34136 Trieste, Italy}

\begin{abstract}
In this work we  study canonical gravity  in finite regions  for which we introduce a generalisation of the  Gibbons-Hawking boundary term including the Immirzi parameter. 
We study the canonical formulation on a spacelike hypersuface with a boundary sphere and  show how the presence of this term  leads to an unprecedented type of  degrees of freedom coming from the restoration of the gauge and diffeomorphism symmetry at the boundary.
 In the presence of a loop quantum gravity state, these boundary degrees of freedom localize along a set of punctures  on the boundary sphere. We demonstrate that these degrees of freedom are effectively described by   auxiliary strings with a 3-dimensional internal target space attached to each puncture. We show that  the string currents  represent the local frame field, that the string angular momenta represent the area flux and that the string stress tensor represents the two dimensional metric on the boundary of the region of interest.
Finally, we show that the commutators of these broken diffeomorphisms charges of quantum geometry  satisfy at each puncture a Virasoro algebra with central charge $c=3$.
This leads to a description of the boundary degrees of freedom in terms of a CFT structure with central charge proportional to the number of loop punctures.
The boundary $SU(2)$ gauge symmetry is recovered via the action of the $U(1)^3$ Kac-Moody generators (associated with the string current) in a way that is the exact analog of an infinite dimensional generalization of the Schwinger spin-representation. We finally show that this symmetry is broken by the presence of background curvature.
\end{abstract}

\maketitle

In loop quantum gravity, quantum geometry is discrete at the fundamental scale and smooth geometry 
at large scales is expected to be a consequence of the coarse graining of Planckian discrete structures. 
In this letter we focus on the presence of boundaries and we show explicitly that a new type of  degrees of freedom are naturally present when one considers  the canonical structure of general relativity on 3-dimensional slices possessing a 2-dimensional boundary.
Remarkably these new excitations, that are initially pure 
quantum-geometry boundary degrees of freedom, behave in some ways as matter degrees of freedom.

The appearance in gauge theories of new degrees of freedom in the presence of a boundary 
has been proposed a long time ago \cite{Benguria:1976in, Balachandran:1991dw, Smolin:1995vq, Carlip:1998wz}, but it is only very recently  that it has been fully  expressed into a coherent picture.
Recently, this question has been revisited with an emphasis on gravity, first in some detail in \cite{Freidel:2015gpa} and taken to a more general level in \cite{Donnelly:2016auv},
 where it was shown  that the boundary degrees of freedom are physical degrees of freedom that restore the gauge symmetry under the presence of boundaries and organise themselves as a representation of a boundary symmetry group. Here we push the analysis further, by carefully studying, in the presence of a generalized Gibbons-Hawking term, the   bulk and boundary components of gauge symmetries. This unravels a Virasoro symmetry as a part of the boundary group.
 
  It is important to understand that  the choice of boundary that decomposes the gravitational system into subsystems corresponds to  
    a choice of observer, and that the degrees of freedom described here are {\it physical} in a precise sense. They represent the set of all possible boundary conditions that need to be included in order to reconstruct the expectation value of all gravity observables, including the non-local ones that involve relations across the boundary. They are needed in the reconstruction of the total Hilbert space in terms of the Hilbert space for the subsystems \cite{Donnelly:2016auv}. They also represent the degrees of freedom that one needs in order to couple the subsystem to  another system in a gauge invariant manner \cite{Rovelli:2013fga}.  These degrees of freedom organize themselves under the representation of a boundary symmetry group. Understanding this symmetry group and its representation at the quantum level is at the core of our paper. This is a  necessary step towards the understanding of the decomposition of gravitational system into subsystems and  the definition of sensible observables in quantum gravity.
    
{At a general level the previous question is completely open at this stage. In this work we make the simplifying assumption that the background geometry in which these degrees of freedom exist is of a loopy nature.
 Let us recall that initially the loop vacuum \cite{Ashtekar:1993wf} was assumed to be such that the fluxes of quantum geometry piercing the boundary vanish outside the flux lines. In the recent years another dual vacuum has been proposed \cite{simplest, Bianchi:2009tj,Freidel:2011ue, Dittrich:2014wpa, Bahr:2015bra} in which 
 it is the gauge curvature that is assumed to vanish outside the punctures. This vacuum possesses a natural geometrical interpretation \cite{Freidel:2010aq, Freidel:2013bfa} and it is in agreement with the spin foam interpretation of quantum gravity \cite{Perez:2012wv}.
 The importance of such state is clear in the context of modelling black hole horizons via the Chern-Simons formulation \cite{Ashtekar:1997yu, Ashtekar:2000eq, Engle:2009vc} and becomes even more explicit in the loop quantum gravity treatment of \cite{Sahlmann:2011xu}.
 We confirm here that it is this new  dual vacuum that admits a boundary interpretation. In this work 
 we assume that the gauge curvature vanishes outside the punctures.}
 
 Under these conditions we are able to find a realisation of the boundary symmetry group that lends itself to quantisation in a direct manner. We find that the boundary degrees of freedom of pure gravity naturally localize
  on the punctures at the intersection of the flux lines with the boundary. This corresponds to a generalisation of  the mechanism first proposed in \cite{Ashtekar:2000eq}. Our implementation is more general since it includes all possible boundary conditions compatible with the extended Gibbons-Hawking term and does not involve an auxiliary (or effective) Chern-Simons theory. Remarkably, we show that each puncture carries a representation of a 3-dimensional Kac-Moody algebra and of a Virasoro algebra of central charge $c=3$. Thus  the boundary degrees of freedom naturally provide a representation of the Virasoro algebra  ${\rm Vir}$ with (an eventually large) central charge $c=3N$, where  $N$ is the number of punctures or elementary geometrical fluxes going through the boundary. 
  
 The conjecture that CFT degrees of freedom should appear
 around loop punctures  was first made in \cite{Smolin:1995vq}. The evidence   that CFT degrees of freedom effectively  appear in the quantisation of loop gravity state  was then first found  in \cite{Freidel:2009nu}. This idea has also  been  explored in the context of loop quantum gravity (LQG) in \cite{Ghosh:2014rra}. Here we give for the first time a full derivation from first principles that 2-dimensional conformal symmetry encodes these so-far elusive excitations. We also find that the nature of these boundary symmetry is related in an intimate way to the presence of the (generalized) Gibbons-Hawking term that renders the action differentiable. 
 
 The paper is organized as follows. In Section \ref{Sact} we introduce the action principle and describe in detail the nature of the generalized Gibbons-Hawking boundary term. We also describe the nature of the variational principle and derive the symplectic structure shown to be preserved by our boundary conditions. In Section \ref{abc} we study the constraints of gravity in the presence of the boundary and compute the constraint algebra. This analysis sets the basis for the rest of the paper and makes apparent the possibility of having extra degrees of freedom at punctures. In Section \ref{Sbound} we define the puncture charges and we show that they satisfy an $U(1)^3$ Kac-Moody algebra. We also construct the associated Virasoro algebra via the standard Sugawara construction.  In Section \ref{FFF} we derive these results again by identifying the degrees of freedom as scalar fields that define a string  with 3-dimensional target space, given by the internal SU$(2)$ representation space. In Section \ref{Kint} we discuss in more detail the relation of our degrees of freedom with the standard ones in LQG, in the case where the curvature takes integer values, and comment on the nature of a CFT/loops duality that had been put forward in the past. The non-integer case is briefly presented in Section \ref{non-integer}.
 In Section \ref{SET} we clarify the link between the present complete formulation and the partial results found in \cite{Freidel:2015gpa} which, {\em a posteriori}, can be regarded as its obvious precursor; we also elucidate the link with the analysis in \cite{Freidel:2009nu}, and we comment on how our finding leads to a new proposal for the loop gravity vacuum state.  We conclude with a discussion of our results in Section \ref{Sdis}.

\section{The action principle} \label{Sact}

We consider a  formulation of 4-dimensional pure gravity on a manifold $M\times \mathbb{R}$, where $M$ represents the 3-dimensional space-like hypersurface, with a boundary 2-sphere $S$. We denote the spacetime boundary $\Delta=S\times \mathbb{R}$.  
We start with an action formulation of vacuum gravity  which includes a boundary term:
\be\label{uno}
S= \frac{1}{\gamma\kappa} \left[ \int\limits_{M\times  \mathbb{R}} \!\! \! E^{IJ}\wedge F_{IJ}(\omega) +\frac12\!\! \int\limits_{ S\times  \mathbb{R}} \!\!\! e_I\wedge \rd_{{B} } e^I\right], 
\ee
where $\kappa=8 \pi G$ and $\gamma $ is the Barbero-Immirzi parameter, $e^I$ is a frame field and $\omega^{IJ}$ is a Lorentz connection, the field $E^{IJ}$ denotes
\be\label{simplicity}
E^{IJ}=   (e^I\wedge e^J)+\gamma *(e^I\wedge e^J),
\ee
while ${B}^{IJ}=\omega^{IJ} + \gamma *\omega^{IJ}$
(where $*$ is the 2-form duality). Even though $B^{IJ}$ in not a Lorentz connection, the quantity $d_Be^I=\rd e^I + B^{IJ} \wedge e_J$ denotes an object with the  tensor structure of a covariant derivative.
Aside from the boundary term (whose geometry we describe below) this action coincides with the usual Holst formulation of first order gravity. 

The boundary densities integrated in \eqref{uno} can be decomposed as the sum of two terms: $*\omega_{IJ}\wedge(e^I\wedge e^J)$ plus  $\gamma^{-1} e_I\wedge d_\omega e^I$.
It is easy to see,  by choosing a Lorentz gauge where one of the tetrad  is fixed to be the normal to the boundary,
that the first component  is simply given by the integral of the  well known Gibbons-Hawking-York boundary density $2\sqrt{h} K$, where $h$ denotes the determinant of the induced metric on the boundary, and $K$ the trace of its extrinsic curvature. The second one is a new addition to the standard boundary term of the metric formulation. The quantity $\gamma^{-1} e_I\wedge d_\omega e^I$ is a natural complement to the Holst term in the bulk action $\gamma^{-1} F_{IJ}(\omega)\wedge e^I\wedge e^J$ that had  already been discussed in \cite{Freidel:2006hv}.  Notice that, as the Holst term, this additional boundary term also vanishes on-shell due to the torsion free condition.  The formal limit  $\gamma=\infty$ corresponds to the usual Cartan-Weyl formulation.

The rationale for the choice of boundary term becomes clear when we look at the equation of motion obtained from variations of the connection $\omega$. On the one hand, the bulk component of this equation is the usual torsion constraint. On the other hand, 
the boundary equation of motion is quite  remarkably given by  the simplicity constraint (\ref{simplicity}).  Interestingly, one can revert the logic and conclude that the simplicity constraint defining the flux-form  follows from demanding the validity of the boundary equation of motion for an arbitrary boundary. This gives an  equivalent point of view and  justification for the choice of boundary term in the action.
 
 The other boundary equation of motion, obtained by varying the boundary co-frame $e$, necessitates extra care.
 As the entire Hamiltonian treatment that will follow makes use of the Ashtekar-Barbero connection formulation, 
 we need to appeal to the availability of the extra structure that allows for the introduction of such variables.
 Such extra structure is the time-gauge---naturally provided by the necessary 3+1 decomposition of the Hamiltonian formulation of gravity via a foliation of spacetime in terms of spacial surfaces $M$---where $e^0$ is chosen so that $e^0=n$, where $n$ is the normal to $M$. We assume that such foliation is available and
  demand the boundary condition $\delta e^0=0$ on $\Delta$. It is very important to understand that we {\it do not} demand, on the other hand, the spacelike frame $e^i$ to be fixed. We let the boundary geometry fluctuate at will, and this turns out to be the source of the boundary degrees of freedom. The set of all admissible boundary frames can be thought of as labelling the set of possible (fluctuating at the quantum level) boundary geometries. These geometries are not arbitrary: they need to satisfy the boundary equation of motion given by the variations of the co-frame at $\Delta$, namely
 \be \label{bconst} \rd_A e^i = 0,\ee
 where $A_a^i$ is the Ashtekar-Barbero connection $A_a^i = \Gamma_a^i +\gamma K_a^i$, with $\Gamma_a^i$  the 3d spin connection and $K^i=\omega^{i0}$ is a one-form related to the extrinsic curvature of $M$\footnote{ One can expand $K^i=K^i{}_{j} e^j$. The symmetric component $K_{(ij)}$ gives the components $K_{ab} e^a_i e^b_j$ of  the extrinsic curvature  tensor to $M$. The skew symmetric part vanish due to the torsion constraint.}.  It is understood that the previous 2-form is pulled back to $\Delta$.
 As a summary, we see that the boundary equations of motion are encoded in the simplicity constraint (\ref{simplicity}) and what we refer to as the (generalized) staticity constraint\footnote{When pulled back to the two sphere $S=\Delta\cap M$ (which will be relevant for the Hamiltonian treatment),  this equation translates into restrictions of the extrinsic curvature. For instance in the gauge $e^3\overset{S}{=} 0$ (in other  words $e^3$ is normal to the sphere $S$) we get  
\be
K^3\wedge e^1\overset{S}{=}0, \ \ \ K^3\wedge e^2\overset{S}{=}0, \ \ \ K^1\wedge e^2-K^2\wedge e^1\overset{S}{=}0.
\ee 
The first two equations imply the staticity constraint $K^3\overset{S}{=}0$ (see \cite{Engle:2010kt}).
The residual non zero components are $K_{AB}$  with $A,B=1,2$. 
The last equation demands that the trace of that tensor vanishes $K_{11}+K_{22}=0$.
This justifies the term `generalized' staticity constraint.} (\ref{bconst}) . 
 
 It can be shown, via the standard covariant phase-space procedure \cite{Ashtekar:1990gc, Lee:1990nz, Witten}, that
 the symplectic form $\Omega= \Omega_M +\Omega_{S^2}$ (a two-form in field space) is, in the present theory,
the sum of a bulk plus a boundary contribution. It reads
\ba \label{symp}
\Omega
&=&\frac{1}{\kappa\gamma }\int_{M} ( \delta A^{i}\wedge \delta
\Sigma^{\va}_{i} ) + \frac{1}{2  \kappa \gamma } \int_{S} (\delta e_i\wedge\delta e^i)\,,
\ea
where $\delta$ denotes the field variation, and the wedge product involves skew-symmetrisation of forms both in space and field space\footnote{ Here we take the convention explained in \cite{Donnelly:2016auv} that $\delta$ is a differential in field space so in particular we have that $\delta^2=0$, that $\delta \phi \wedge \delta \psi = -   \delta \psi \wedge \delta \phi$ and the product is such that $(\delta \phi\wedge \delta \psi)(V,W) = \delta_V \phi \delta_W \psi  -\delta_W \phi \delta_V \psi $ for fields variations $(V,W)$.}.
 The bulk configuration variable is an SU$(2)$ connection $A^i$ and the variable conjugate to this connection is  the flux-form $\Sigma_i$, a Lie algebra valued two-form. 
From this we read the Poisson bracket of the bulk phase-space, which is given as usual by 
$\{ A_a^i(x),\Sigma_{bc}^j(y) \} = \kappa \gamma \delta^{ij}\epsilon_{abc}  \delta^3(x,y)$.
We can also read the  boundary phase-space structure \cite{Freidel:2015gpa}
\be\label{cr}
\{e^i_a(x), e^j_b(y)\}= { \kappa \gamma}\, \delta^{ij} \epsilon_{ab}\delta^2(x,y)\,.
\ee

In summary, we have that 
the bulk fields are given by an SU$(2)$ valued  flux-2-form $\Sigma_i$ and an SU$(2)$ valued connection $A^i$ satisfying the scalar constraint
$e_i\wedge F^i(A)+\cdots=0$ (the dots here refer to a term involving the extrisic curvature and proportional to $(\gamma^2+1)$), the Gauss law $\rd_A \Sigma_i=0$, and the diffeomorphism constraint  $ F^i(A)\wedge [\varphi, e]_i=0$, with   $\hat \varphi$
a  vector field\footnote{ The diffeomorphism constraint is usually written  in terms of a vector $\hat\varphi=\varphi^a\pa_a$ tangent to $M$ as $[\hat\varphi\righthalfcup F^i(A)]\wedge \Sigma_i=0$. Using that $[\hat\varphi \righthalfcup F]\wedge \Sigma + F\wedge[\hat\varphi \righthalfcup \Sigma]=0$ and defining $\varphi^i\equiv 
[\hat\varphi\righthalfcup e^i]$ we obtain the expression in the main text. } tangent to the slice $M$.
Bulk fields $(\Sigma_i,A^i)$ can be seen as background fields that commute with the boundary field $e_i$. 
Finally, the preservation of the gauge and diffeomorphism symmetry in the presence of the boundary imposes the 
validity of additional boundary constraints. In agreement with \cite{Freidel:2015gpa}, we find that the boundary diffeomorphism constraint is associated with the generalised staticity constraint (\ref{bconst}), while the boundary gauge constraint is given by\footnote {The SU$(2)$  bracket is taken to be $[X,Y]_i=\epsilon_{ijk}X^j Y^k$.}
\be\label{bbconst}
\Sigma_i =\frac12[e,e]_i.
\ee
The previous constraint is the boundary Gauss law, a boundary condition which identifies the bulk flux-form with its boundary counterpart. This is essentially the simplicity constraint (\ref{simplicity}) pulled back on a slice and written in terms of canonical variables. In this way the simplicity constraint is here the  condition enabling the preservation of SU$(2)$ symmetry in the presence of a boundary. 
It is important to understand that one treats $\Sigma$ and $e$ as {\it independent} fields. In particular $e$ commutes initially with all the bulk fields $A$ and $\Sigma$ as follows from (\ref{symp}), and  $\Sigma$ commutes with itself.

Thus, the relationship between the bulk variables $(\Sigma,A)$ and the boundary variables $e$ is encoded into two constraints: The boundary Gauss constraint (\ref{bbconst}) which relates the pull-back of $\Sigma $ on $S$ with $e$ and the staticity constraint (\ref{bconst}) that relates the pull-back of $A$ with  $e$ on the boundary $S$.
In particular we see that it is the boundary Gauss constraint that implies  that the boundary fluxes do not commute, while their bulk version does.  

Our goal now is to study the quantisation of this boundary system in the presence of the background fields.
In order to understand this point it is crucial  to appreciate that the pull-back 
$\Sigma^i_{z\bar{z}}$ of the flux-form on $S$  { and} the  components of the connection $(A^j_z,A_{\bar{z}}^j)$ tangential to $S$ {\it commute} with each other. Here {we have denoted by $(z,\bar{z})$ the complex directions tangential to $S$}. From now on we assume that a particular complex structure on the sphere
has been chosen.
Because these components commute, we can a priori fix them to any value on the boundary and study the boundary Hilbert space in the presence of these boundary fields.
Once  we implement the simplicity constraint (\ref{bbconst}) we determine the value of the boundary flux $\Sigma^i_{z\bar{z}}$ in terms of the boundary frames $(e_z^i,e_{\bar z}^i)$. We are still free to chose at will the value of the boundary connection $(A_z^i,A_{\bar z}^i)$ and study the implications of the staticity constraint.
 As we have seen, this boundary phase-space represents the set of admissible boundary conditions encoded into the frames  $e^i$.
 
{
We are going to make the key assumption---motivated by the new developments \cite{Bianchi:2009tj,Freidel:2011ue, Dittrich:2014wpa, Bahr:2015bra} in quantum gravity---that the tangential curvature of $A$ vanishes everywhere on the sphere except at the location of a given set of  $N$ punctures $P\equiv \{x_p\in S| p=1,\cdots , N\}$ defined by the endpoint of spin-network links. 
This is possible due to the fact that the tangential connection commutes with the flux $\Sigma$ which is fixed by the simplicity constraint (\ref{bbconst}). 
Let us recall that  given a disk $D$ embedded in $S$ 
we can define the flux and holonomy associated with $D$ as $\Sigma_D^i \equiv \int_D \Sigma^i $ and $g_D \equiv P\exp\oint_{\pa D} A$ respectively.  Our assumption therefore means that we impose the conditions 
\be
 g_D=0,
\ee
for disks $D$ in $\overline D=S\backslash P$. 
We do not impose any restrictions on  the value of the fluxes outside the punctures since this value is now controlled by the boundary condition  (\ref{bbconst}). 
In the following we will impose the  condition
on the  curvature at the puncture by demanding that  we have  $ g_{D_p} = \exp{2\pi K_p}$, for  a disk $D_p$ around the puncture $x_p$.  
The location of the punctures $x_p$ and the SU$(2)$ Lie algebra\footnote{ We parametrize SU$(2)$ Lie algebra elements by anti-hermitian operators.} elements  $K_p^i$ parametrize the   background curvature of the boundary. 
In other words we impose that  the curvature's connection  is such that \be\label{curv} F^i(A)(x) = 2\pi \sum_p K_p^i \delta^{(2)}(x,x_p).
\ee 
This condition is  natural from the point of view of the bulk constraints since the vanishing of the curvature also implies the vanishing of the  bulk diffeomorphism constraint, which is a condition on curvature. Dual spin network  links piercing the boundary are thus labelled by $K_p$, which expresses the fact that, from the perspective of $S$, they are a source of tangential curvature.
}

\section{Algebra of boundary constraints}\label{abc}
We now assume that 
the 2-sphere $S = \bar{D} \cup_p D_p$ can be decomposed into a union of infinitesimal disks $D_p$ surrounding the puncture $p$  and its complement denoted $\overline{D}$.
The two generators  associated with the two constraints \eqref{bconst} and  (\ref{bbconst}) are obtained from the symplectic structure through\footnote{The Poisson bracket is related to the symplectic structure via $\{F,G\}=\Omega(\delta_F,\delta_G)$ where $\delta_F$ is the Hamiltonian variation generated by $F$, $\Omega(\delta_F,\delta)= \delta F$.  
 }
\ba
\Omega(\delta_\alpha,\delta) = \delta G_D(\alpha),\qquad  \Omega(\delta_\varphi,\delta) = \delta S_D(\varphi),
\ea 
 and they read
 \ba\label{ggaauu}
&& G_D(\alpha) \equiv \frac{1}{\kappa\gamma} \left(\frac12\int_{D} \alpha^i [e,e]_i - \int_M\rd_A\alpha^i \wedge \Sigma_i\right) , 
 \\ && S_D(\varphi) \equiv \frac{1}{\kappa\gamma}\left(\int_{ D}  \rd_A\varphi^i   e_i +  \int_M F_i(A)\wedge[e,\varphi]^i\right)\,.\n
\ea

We see that the constraint $G_D(\alpha)$ is a boundary extension of the Gauss constraint generating gauge transformations for the bulk variables.
By integrating by parts the bulk term we see that it imposes the gauss Law $\rd_A\Sigma^i=0$ and the boundary simplicity constraint (\ref{bbconst}).
It is also  the generator of internal rotations $\delta_\alpha e_i =[\alpha,e]_i$ for the boundary variables. 
The subscript $D$ refers to the condition that the parameter $\alpha$  vanishes outside of $D$ and is extended inside $M$. 
The constraint $S_{D}(\varphi)$ is a boundary extension of the diffeomorphism constraint  for the bulk variables when 
$\varphi^i =\hat{\varphi}^a e_a^i$ for a vector field $\hat\varphi$ tangent to $M$.
There is a subtlety with the `staticity' constraint $S_D(\varphi)$: it is differentiable {\it only} in the form written here. When $\pa D=\emptyset $ we can integrate by part and the boundary term is proportional to $\int_{ D}  \varphi^i   \rd_A e_i$, which imposes the staticity constraint.
In the general case where $\varphi$ does not necessarily vanish on $\pa D$ we need to add a corner term to the staticity constraint and write it as $\int_{ D}  \rd_A \varphi^i    e_i$ so that its variation is well defined.
 Computing this variation we conclude that  $S_D(\varphi)$ generates, diffeomorphism in the bulk and for the boundary variables, the transformations 
$\delta_\varphi e^i= \rd_A \varphi^i$ with $\varphi$ supported on $D$.

 By a straightforward but lengthy calculation, it can be verified that the constraint generators satisfy the following algebra 
\ba\label{Algebra}
\{G_D(\alpha),G_D(\beta)\}&= G_D([\alpha,\beta]),\qquad\qquad
\qquad\quad
\\
\{ G_D(\alpha),S_D(\varphi)\}&= \int_{\pa D} ( [\varphi,\alpha]_i  e^i)
+S_D([\alpha,\varphi]),\n\\
\{S_D(\varphi), S_D(\varphi')\} &\hat{=} \int_{\pa D}(\varphi^i  \rd_A\varphi'_i) - \int_D F^i[ \varphi,\varphi']_i, \n
\ea
where the $\hat{=}$ means that we have imposed, in the last equality, the constraints:  $G_D= S_D=0$.
The  structure of this algebra is one of the key result of this paper and of central importance.
One sees that in general, the boundary diffeomorphism  algebra is second class with the appearance of central extension terms supported  on the boundary of the domain $\pa D$. In the case when $F\neq 0$ there exists additional second class constraints supported entirely on the domain $D$. 
 We are witnessing  here the  mechanism behind the generation of degrees of freedom: At the location of the punctures where boundaries appear (and where $F\neq0$)  the constraints become second class and, since a first class constraint removes two degrees of freedom while a second class only one, this means that at the punctures some of the previously gauge degree of freedoms become now physical.
This analysis is valid only classically. What we are going to see at the quantum level is that this phenomenon is accentuated by the appearance of anomalies in the diffeomorphism algebra.

\section{Boundary charges} \label{Sbound}
Here we show explicitly how a $U(1)^3$ Kac-Moody algebra---whose generators are closely related to those of singular diffeomorphims at punctures---is associated with punctures.
In order to do so we introduce boundary charges 
\be\label{Q}
Q_D(\varphi)\equiv\frac1{\sqrt{2\pi \kappa \gamma}}  \int_D \rd_A \varphi^i \wedge e_i,
\ee
where $\varphi=\varphi^i\tau_i$ is an su$(2)$ valued field who enters only through its covariant derivative. Hence, without loss of generality  we assume that it vanishes at the puncture $\varphi(p)=0$. 
After integration by parts this becomes
\be\label{discon}
Q_D(\varphi) = \frac1{\sqrt{2\pi \kappa \gamma}}  \oint_{\partial D} \varphi^i e_i - \frac1{\sqrt{2\pi \kappa \gamma}} \int_D \varphi^i(\rd_A e_i).
\ee
We therefore see that the  charge $Q_D(\varphi)$ depends only on the boundary value of the field, once the staticity constraint \eqref{bconst} is imposed. 
Moreover from (\ref{discon}) we see that the canonical charges $Q_{\overline{D}}(\varphi)$ can be decomposed as a sum around each puncture $ Q_{\overline{D}}(\varphi) \hat{=}  -\sum_p Q_p(\varphi)$, 
where the hatted equality means that we have imposed the staticity constraint.
 Concretely, when we focus on a single puncture $p$, its contribution can be expressed as a circle integral
\be\la{charges}
Q_p(\varphi) \hat{=} \frac1{\sqrt{2\pi \kappa \gamma}} \oint_{C_p} \varphi^i e_i,\qquad
\ee
where $C_p$ is an infinitesimal circle  around the given puncture with the orientation induced by that on $D_p$.
From the  expression (\ref{Q}) and  using \eqref{cr}, we can directly compute the commutator of the charges 
$Q_p(\varphi)$.  We get 
\ba\label{15}
&& \{ Q_p(\varphi),Q_{p'}(\psi) \}
 = \n \\ &&=
 {\delta_{pp'}}\left(   K_p^i [\varphi, \psi]_i(p) +\frac1{2\pi}
 \oint_{C_p} \varphi^i \rd_A \psi_i \right).
\ea
Since the fields are assumed to vanish at $p$ 
the first term vanishes.
Next, in order to evaluate the integral we can chose a gauge around punctures. More precisely, the condition $F^i(A) = K^i \delta(x)$ can be solved in the neighborhood of the puncture in terms of 
$A = (g^{-1}Kg) \rd \theta   + g^{-1}\rd g $, 
where we chose polar coordinates $(r, \theta)$ around the puncture
and denoted $g$  a group element which is the identity at the puncture. 
We can fix the gauge $g=1$. In this gauge the gauge field is constant with $A= K_p  \rd \theta$, the fields are periodic and   we discover a twisted $U(1)^3$ Kac-Moody algebra per puncture, namely
\be\label{algebra1}
\{ {Q}_p(\varphi),{Q}_{p'}(\psi) \} = \, \frac{\delta_{pp'}}{2\pi}\oint_{C_p} (\varphi^i \rd \psi_i -K^i_p [\varphi,\psi]_i\rd \theta).
\ee 
An important property of the $Q_p(\varphi)$ is that even though they are defined by an integral involving the bulk of $D_p$, their commutation relations depends only on the values of the smearing field $\varphi$ on the boundary of $D_p$.

By defining the modes of the charges \eqref{charges} as\footnote{We work in an anti-hermitian basis $\tau^i$ where $[\tau^i,\tau^j]=\epsilon^{ijk}\tau_k$.}
 \be\la{Qmodes}
Q_n^j\equiv Q(\tau^je^{i \theta n})\,,
\ee
where $\theta$ is an angular coordinate around $C_p$, and $\tau^i$ are $\su(2)$ basis vectors.
From \eqref{algebra1} we get
that the algebra becomes \ba\label{algebra2}
\{Q^i_n, Q^j_m\}
&=& \delta_{n+m} (im \delta^{ij} +\tau^i[K,\tau^j])\n\\
&=& -i\left(n\delta^{ij}+ K^{ij}\right) \delta_{ n+m}\,.  
\ea
where we have defined\footnote{In other words $[K,\varphi]^i=- 
 iK^{ij}\varphi_j$.} $K^{ij}:= i\epsilon^{ikj} K_k$. 
In the case where the curvature vanishes  we simply get 
$
\{Q^i_n, Q^j_m\}
=-in \delta^{ij}\delta_{ n+m}
$,
which corresponds to three Abelian Kac-Moody algebras, each with central extension equal to $1$.  In the presence of curvature, we obtained a three dimensional abelian Kac-Moody algebra twisted by $K$. 
It will be convenient to work in a basis $\tau^a=(\tau^3,\tau^+,\tau^-)$ \footnote{We define $\tau^\pm=(\tau^1\mp i\tau^2)/\sqrt{2}$, $[\tau^3,\tau^\pm]= \pm i \tau^\pm$, $[\tau^+,\tau^-]= i\tau^3 $.}  where 
 \be K= k\tau_3,
 \ee 
 and $K^{a\bar b}$ is diagonal.
In this basis the non trivial commutators of  the twisted algebra are given by
\ba
\{Q^3_n,Q^3_m\}&=& -in\delta_{n+m},\n \\ 
\{Q^+_n,Q_m^-\}&=&-i(n+k) \delta_{n+m}.
\ea
We will mostly work in this complex diagonal basis in the following and we will denote 
$k^a:=(0,+k,-k)$, where $(a=3,+,-)$ and $\bar{a}=(3,-,+)$ denote the conjugate basis; in this basis the metric is $\delta_{a\bar{b}}$. It will be relevant that $\sum_a k_a=0$.  The Poisson bracket can be promoted to commutator $[\cdot,\cdot]=i\{\cdot,\cdot\}$. The twisted Kac-Moody algebra can then be written compactly as 
\be
[Q^a_n,Q_m^{b}]= \delta^{a\bar{b}}(n+ k^a) \delta_{n+m}. 
\ee
We will restrict in the following to $k$ being in $\mathbb{Z}/N$ for some integer $N$ (such restriction will be clarified below). We will also see that the theory associated with $k$ and with $k+1$ are in fact {\it equivalent}.
This equivalence corresponds to the fact that at the quantum level the connection is compactified. A fact that usually follows from loopy assumptions but here is derived completely naturally in the continuum. 
The appearance of $k^a$, in the previous equation will be rederived in Section \ref{FFF}.

 \subsection{Sugawara construction}
 Up to now we have focused on the Kac-Moody charges conjugated to an internal vector.
 It is however also interesting to focus on the generators of boundary diffeomorphisms that generate covariant diffeomorphism along a vector field $v^a\pa_a$ tangent to $S$.
 The covariant version of the Lie derivative is defined to be
 \be
 L_{v} e^i :=  v\righthalfcup \rd_{A}e^i+\rd_{A}(v\righthalfcup e^i),
 \ee
 and the corresponding boundary charge is 
 \be
 T_D =\frac{1}{2\kappa \gamma} \int_D L_ve^i \wedge  e_i\,,
 \ee
 as it can be checked from the relation $\Omega_D(L_v,\delta)= \delta T_D$ for variations that preserve $A$. When the staticity constraint \eqref{bconst} is satisfied, the previous expression can  be simply written as 
 \be\la{TD}
  T_{D_p}(v) =\frac{1}{2\kappa \gamma} \oint_{\pa D_p} (v\righthalfcup e^i) e_i. 
 \ee
We can introduce the modes $L_n^{(p)}\equiv T_{D_p}(\exp(i\theta n) \partial_\theta)$, explicitly 
\be\la{Ltheta}
L_n = \frac{1}{2\pi}\oint e^{i\theta n}\, T_{\theta\theta}\, \rd \theta\,,
\ee 
 where the integrand is the $\theta\theta$ component of the energy-momentum tensor
 \be
 T_{\theta\theta}=\frac{\pi e_\theta^i e_{\theta i}}{\kappa \gamma}\,.
 \ee
 It is straightforward to show that the modes \eqref{Ltheta} can be obtained from the Kac-Moody modes $Q^a_n$, defined in \eqref{Qmodes}, through the Sugawara construction \cite{DiFrancesco:1997nk} and that they satisfy a Virasoro algebra with central charge $c=3$. More precisely, following the classical analog of  the standard Sugawara construction applied to the Kac-Moody currents one defines
 \be\la{Sug}
L_n = \frac12 \sum_a\sum_{ m\in\,\Z} Q_m^a Q_{n-m}^{\bar{a}}\,.
\ee
 Equivalence between \eqref{Ltheta} and \eqref{Sug} can be easily checked by means of \eqref{Qmodes}.
At the quantum level we introduce
\ba\la{L}
&&  L_{n} = \frac12 \sum_a\sum_{m\in\,\Z}\! : Q^a_{m}  Q^{\bar a}_{n-m}:\,,
\ea
 where we omit hats to denote quantum operators as the context 
 clarifies their quantum nature, and  $:\ \ :$ stands for the normal ordering defined by
\be
:{Q}^a_{n}{Q}^b_{m}\!:\,=\begin{cases}
         Q^b_{m} Q^a_{n}\quad\text{if}\;n+k^a>0\\
          Q^a_{n} Q^b_{m}\quad\text{if}\;n+k^a\leq0\,.\end{cases}  
\ee
From the quantisation of the algebra \eqref{algebra2}  we get
\ba\la{QQ}
[{Q}^a_{ n}, {Q}^{ b}_{ m}] 
&=&{\delta^{a \bar b}}  (n+k^a) \delta_{n+m}\,.
\ea
It follows from standard considerations \cite{Borisov:1997nc, Evslin:1999qb}, and it can be checked through a straightforward but lengthy calculation, that the generators $ L_{n}$  satisfy the Virasoro algebra
\ba\la{Vir}
[ L_{n},  L_{m}]=(n-m)  L_{n+m} + \frac{c}{12}n(n^2-1)\delta_{n+m,0}\, ,
\ea
with $c=3$. 
It is also convenient to write the algebra of the Kac-Moody with the Virasoro modes
\be\la{LQ-1}
[ L_{n},  Q^a_{m}]=-(m+k^a) Q^a_{n+m}\,.
\ee
This shows that the currents are  primary fields\footnote{ Lets recall that an  untwisted primary field of weight $\Delta$ satisfies $[L_n,O^\Delta_m]= (n(\Delta-1)-m)O^\Delta_{m+n}$.}  of weight $1$ twisted by $k$.
\subsection{Intertwinner}
{
At this point we can also recover the $SU(2)$ local symmetry algebra
generated by the {\em l.h.s.} of \eqref{bbconst}. Here we simply give an algebraic construction
while we postpone the derivation of its relation with \eqref{bbconst} to Section \ref{Sloop}. 
Of course we have seen in eq. (\ref{Algebra}) that the generator $G_D(\alpha)$ associated with  a region $D$ generates SU$(2)$ transformations.
However in  section \ref{abc} we included in the transformations the variation of the background fields, namely the connection $A$. What we are now looking for is a transformation that affects only the boundary modes $Q_n$ while leaving the background fields invariant. Such transformations are {\it symmetries} of the boundary theory that interchange different boundary conditions without affecting the solutions in the bulk. 

The properties of these symmetries thus depend  on the bulk variables.
In the previous section we have seen that the tangential diffeomorphisms forming a Virasoro algebra act on the boundary variables $Q_n^a$. 
If we assume first that the curvature vanishes at the punctures then we find that 
there exists also an SU$(2)$ generator acting on the boundary variable.
}
Concretely,  when $k^a=0$, we define
\ba\la{schw}
 M^i
&=&\epsilon^i{}_{jk} \sum_{n\neq 0} \frac{Q^j_nQ^k_{-n}}{2 n}\,.
\ea
 It is then straightforward to verify using \eqref{QQ} and the vanishing of $k^a$ that 
\be\label{su2-1}
[M^i,M^j]=\epsilon^{ij}\!_k\, M^k\,.
\ee
Notice that the expression \eqref{schw} can be viewed as an infinite dimensional analog of the
Schwinger representation of the generators of rotations. This provides a new representation of the $\su(2)$ Lie algebra generators in terms of the $U(1)^3$ Kac-Moody ones. 
These generators represent the quanta of Flux at each puncture given by 
\be
M^a_p= \int_{D_p} \Sigma^a.
\ee
As such they are therefore the generalisation of the loopy flux variable.
We will show later that these generators are 
associated with the angular momentum of the puncture in their stringy interpretation. 
Finally, from \eqref{LQ-1} we get that 
\be
[L_n, M^i]=0.
\ee
Therefore in the case $K=0$
we are exhibiting, in an explicit way, the existence of a Virasoro algebra with central charge $c=3$ at each individual puncture
associated to residual diffeomorphisms times the well known $\su(2)$ local algebra (\ref{su2-1}) of LQG that is preserved by the 
Virasoro generators.  The relationship of these generators and the presence of CFT degrees of freedom is our main result.

When $K\neq 0$ (and not integer) the SU$(2)$ symmetry is broken down to a $U(1)$ symmetry that preserves the connection. The unbroken generator of  
$U(1)$ symmetry is given by 
\be
M^3 = -i \sum_{n\in \mathbb{Z}} \frac{Q_n^+ Q_{-n}^-}{n+k}.
\ee
In the case when $K$ takes integer values, there still is a residual SU$(2)$ symmetry that leaves the background connection invariant and the angular momentum generators satisfy an $\su(2)$ algebra. The explicit construction will be presented in Section \ref{Sloop}.

{
To summarise, we have seen that each puncture $p$ carries a representation of the product of Virasoro $L_n^{p}$ times an SU$(2)$ or U$(1)$ generated by $M^{p}$, depending on wether there the curvature takes integer values or not.
This can be understood as a thickening of the spin network links in terms of 
cylinders in the spirit of \cite{Smolin:1995vq, Haggard:2014xoa}  (see figure \ref{Figure}).
\begin{figure}[h]
\centerline{ \(
\begin{array}{c}
\includegraphics[height=3.3cm]{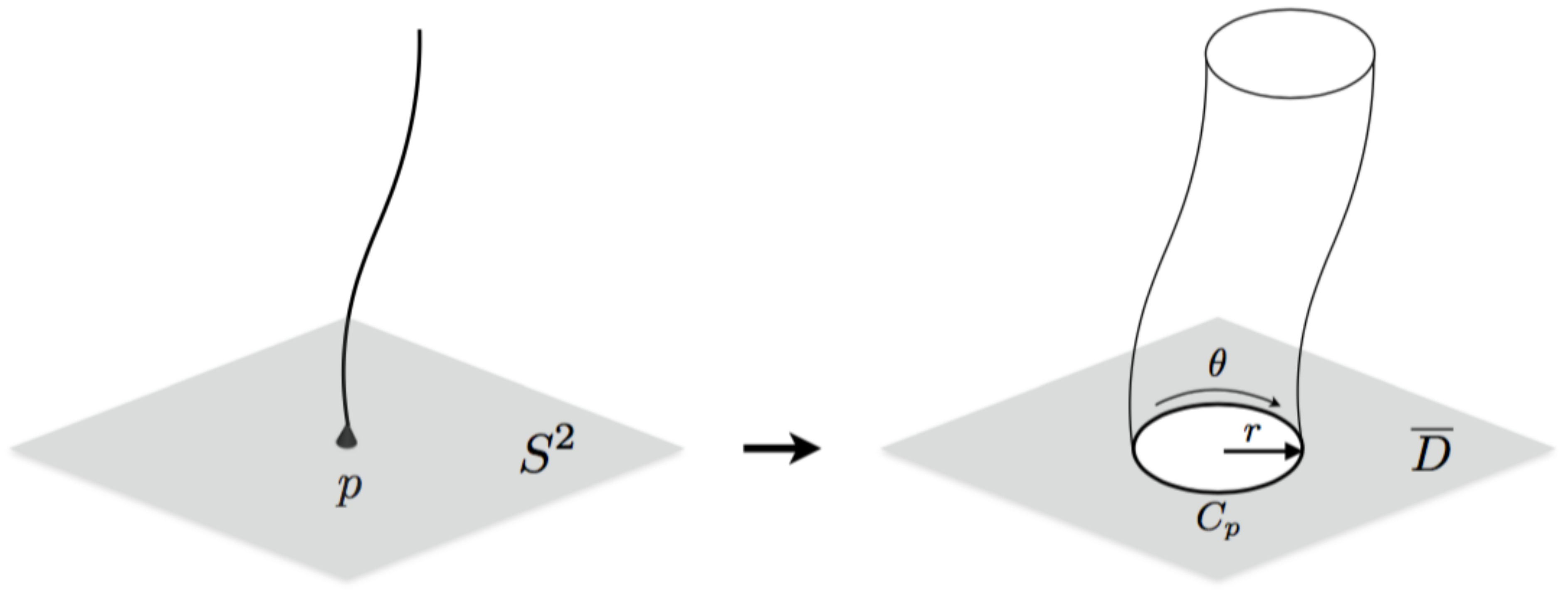}
\end{array}\) } \caption{The thickening of dual spin-network links into Virasoro spin-tubes.}
\label{Figure}
\end{figure}

It is interesting to note that in general the naive conservation law for these generators are not satisfied. Indeed we have that 
\be\la{lala}
\sum_p  L_n^{p} = - L^{\overline D}_n,\qquad 
\sum_p  M^{p} = -M^{\overline D}.
\ee
The violation of the closure constraint is encoded into the value of the generator $M^{\overline D}$ with support outside the puncture. 
There is a  way to read the previous equations in which $L^p$ is a symmetry generator associated to each puncture, coming from the intersection of the dual spin network edge with $S$,
and $ L^{\overline D}$ represents the {\it intertwiner}---or vertex in the loop language---linking together  the different  spin network edges.

In usual Loop gravity it is assumed that the flux   vanishes outside the punctures. This would translate into the statement that $M^{\overline D}=0$, or in other words that the intertwiner is assumed to be uncharged. Here we see that we can relax this hypothesis. This fact is important since it was understood by Livine \cite{Livine:2013gna} that, under coarse graining, curvature is generated and this implies that the naive closure constraint expressed as $M^{\overline D}=0$ is then violated. This shows vividly that the original loop vacuum containing vanishing flux is not stable under renormalisation \cite{Charles:2016xwc}. In our construction the fact that the generators $M^{\overline D}$ do not vanish but take values determined by the puncture data suggests strongly that we have a description which the vacuum is stable under coarse graining.
}
In order to understand the nature of the symmetry `intertwiners' 
$(L^{\overline D}_n,Q_n^{\overline D a}, M^{\overline D})$ associated with the complementary region, we now study the solution space outside the punctures
and we establish that the intertwiner is a 3-dimensional auxiliary string.

\section{The string target space}\label{FFF}

In this section we will recover previous results in an alternative way. Concretely, we will resolve the staticity constraint \eqref{bconst} outside punctures via a gauge fixing. This will allow us to establish a direct link between the algebraic structures found  and the currents 
arising from a 2-dimensional CFT on the boundary, and thus provide a stringy interpretation of the new degrees of freedom.

Let us first  understand the nature of the first class constraints outside the punctures. On $\bar{D}$ we have 
\be
\rd_A e^i =0\,,\qquad F(A)=0\,.
\ee
The zero curvature equation can be easily solved by 
$ A= g^{-1} \rd g $, while the normalisation condition on the curvature imposes that around each puncture $p$, blown up to the circle $C_p$ of radius $r$, we have that the group element is quasi-periodic: 
\be
g(z_p + r e^{2i\pi} )=  e^{2\pi K_p} g(z_p +r).
\ee
Using this group element we can redefine  the frame and internal  vector as $e^i= (g^{-1}\hat{e}^ig)$ and $\varphi^i = (g^{-1} \hat{\varphi}^i g)$.
The unhatted quantities are periodic while the hatted ones are only quasi-periodic.  In the hatted frame the connection $A$ vanishes. The only effect to  keep in mind is the quasi-periodicity of the fields around the punctures $ \hat{e}(z_p + r e^{2i\pi} )=  e^{2\pi K_p} \hat{e}(z_p +r)$.

We assume that this gauge is chosen and we can therefore neglect the connection $A$ in the following equations, as it is assumed to be flat on $\bar{D}$.
As previously stated, the Hamiltonian action  of \eqref{Q} on $e^i$
generates the translational gauge transformation which we can now write as 
\be
\hat{e}^i \to \hat{e}^i+ \rd\hat{\varphi}^i. 
\ee
We concentrate on those transformations with $\hat\varphi^i=0$ on the boundaries $\partial \bar{D} = \cup_p C_p$. 
These correspond to the transformations generated by  the
staticity constraint \eqref{bconst} $S_{\bar{D}}(\varphi)$ and are linked via $S_{\bar{D}}(\varphi\righthalfcup e^i )$ to the tangent bulk diffeomorphism 
that are trivial at punctures but otherwise move them around.  
We can define a natural gauge fixing for them by choosing a background metric $\eta_{ab}$
and imposing the gauge condition 
\be\label{gg}
G^i\equiv \eta^{ab}\partial_a \hat{e}_b^i =0\,.
\ee
That this  is a good gauge fixing condition for the above subclass of transformations follows from the fact that
the only solution to the equation \be
0=\delta_\varphi G=\{G, D_S(\hat\varphi)\}=\Delta \hat\varphi^i\,,
\ee
satisfying the boundary condition $\hat \varphi^i=0$ on the punctures $C_p$,  
is the trivial solution $\hat\varphi^i=0$ everywhere on $\overline D$. Only trivial transformations generated by 
\eqref{bconst} leave the gauge condition invariant.

Notice now that the general solution of the staticity constraint $d\hat{e}^i=0$ can be written as
\ba\la{ephi22-bis}
\hat{e}^i=\sqrt{\frac{\kappa \gamma}{2\pi}} \,\rd X^i,
\ea
where $X^i$ are scalar fields whose normalization is chosen for later convenience.  After plugging \eqref{ephi22-bis} into the gauge condition $G^i=\rd *\hat{e}^i=0$  we obtain 
\be\label{lapla}
\Delta X^i=0\,.
\ee
In fact the introduction of $\eta_{ab}$ in \eqref{gg} amounts to a choice of a complex structure on $S$: concretely 
it defines complex coordinates $z$ such that $\eta=\rd z\rd\bar z$. 
The result of introducing the gauge fixing of the staticity constraint and solving these 
requires the parametrization of the remaining degrees of freedom in terms of holomorphic and anti-holomorphic solutions of
\eqref{lapla}. 
In terms of complex variables, the gauge fixing takes the form
\be
  G\equiv\partial_z \hat{e}^i_{\bar{z}}+\partial_{\bar{z}} \hat{e}^i_z=0,
\ee
and the staticity constraint $\rd_Ae^i=0$ becomes
\be
\partial_{{\bar z}} \hat{e}^i_z-\partial_{\bar z} \hat{e}^i_{{z}}=0\,.
\ee
The solution of the staticity constraint and the gauge fixing is therefore given in terms of the value of three  scalar fields $X^i$ which are solution of  the Laplace equation on the sphere
\be
\Delta X^i =0\,.  
\ee
 If we  use on $\bar{D}$  conformal coordinates $(z,\bar{z})$, this solution can be factorised into the sum of a  left and right movers:
$X^i=X^i_+(z) + X^i_-(\bar{z}) $, where $\pa_{\bar{z}} X^i_+ = 0$, $\pa_{{z}} X^i_-=0$.
The value of such a field on $\bar{D}$ is entirely determined by its value on the circles $C_p$ that compose the boundary $\partial \overline D$.
The frame fields are proportional to the conserved  currents:
$ \hat e^i_z =\sqrt{\frac{\kappa \gamma}{2\pi}}\, J^i$ and $ \hat e^i_{\bar{z}} =\sqrt{\frac{\kappa \gamma}{2\pi}}\, \bar{J}$, where 
\be \la{J-X}
 J^i:=\pa_z X^i, \qquad  \bar{J}^i:= \pa_{\bar{z}} X^i.
\ee
These two copies are not independent as they are linked together by the reality condition 
\be\label{rc}
(\hat{e}^a_z)^* = \hat{e}^{\bar a}_{\bar z}\,. 
\ee
The equations of motion implies that $J^i$ is a holomorphic current while $\bar{J}^i$ is an anti-holomorphic one, and they satisfy around the puncture $p$ the quasi-periodic conditions 
$J(z_pe^{2i\pi}) = e^{ 2\pi K_p}J(z_p)e^{-2\pi K_p}$ and 
$\bar J(\bar{z}_p e^{2i\pi}) = e^{-2\pi K_p} \bar{J}(\bar z_p)e^{2\pi K_p}$.
As we have already seen it is convenient to work in an internal  frame where $K_p= k_p\tau^3$ and work in complex basis $\tau^a=(\tau^3,\tau^\pm)$, which diagonalises the adjoint action, instead of the real basis $\tau^i$. In the complex basis $(\tau^a)^\dagger = - \tau^{\bar a}$, $\tau^{\bar a}= \tau_a$ and 
$[K_p, \tau_a]= -i k^a_p \tau_a$  
with $k^{\bar a}_p = -k^a_p$ and the currents satisfy the quasi-periodicity condition
\be\label{QP}
 J^a(z_pe^{2i\pi}) = e^{-2i\pi k^a_p}\!J^a(z_p), \quad
 \bar J^{ a}(\bar{z}_p e^{2i\pi}) = e^{2i\pi k^a_p}\! \bar{J}^{ a}(\bar z_p).\nonumber
\ee

We can now pull back the symplectic structure  \eqref{symp} to the solutions of $G^i=d\hat{e}^i=0$ parametrized by the 
scalar fields $X^i$. Concretely, starting from \eqref{symp} 
and using (\ref{ephi22-bis}), we have $\Omega_{\bar{D}}=-\sum_p \Omega_p$ where 
\ba\label{symppp}
&&  \Omega_p=\frac{1}{2  \kappa \gamma } \int_{D_p} \delta e_a\wedge\delta e^{ a} 
= \frac{1}{4\pi} \int_{C_p}  \delta X_a \, \rd \delta  X ^{ a} . 
\ea
It is important to note that the integrand in (\ref{symppp}) is periodic  
because it contains the contraction of two fields  and can be written in the original or the hatted quasi-periodic frame.

We are now ready to rederive \eqref{algebra2} in terms of the current algebra. The idea is to express the symplectic form (\ref{symppp}) in terms of the modes of the individual currents $J^i$ and ${\bar J}^i$. 
We will from now on always work around a given puncture and therefore drop the label $p$. We will also work in the complex frame introduced above that diagonalises $K^{ij}$. 
First lets recall that the holomorphicity of the currents and quasi-periodicity condition (\ref{QP})
imply that the currents admit the expansion:
\be\label{J}
z J^a(z) = \sum_{n\in \mathbb{Z}} J^a_n z^{-n-k^a},
\qquad 
\bar z \bar{J}^{{a}}(\bar z) =\sum_{n \in \mathbb{Z}} \bar{J}^{{a}}_n \bar{z}^{-n+ k^a}.
\ee 
Note that in order to make sense of such expansions we have to restrict to curvatures that satisfy the condition  that $k^a \in \mathbb{Z}/N$ for some integer $N$
\footnote{This is a pre-quantization condition on the distributional curvature at punctures of a form that is familiar to
other more restrictive formulations of boundary conditions (see \cite{G.:2015sda} and references therein).}.
The reality condition (\ref{rc}) gives the identification
\be\label{real}
(J^a_{n})^\dagger  = \bar{J}_{n}^{\bar a}\,.
\ee

Before proceeding let us point out that the reality conditions (\ref{real}) are indeed very   
 different from the 
usual ones appearing for instance in string theory.
The difference comes from the fact that usually one quantises  a scalar field $Y$, solution of a {\it Lorentzian} wave equation $\square Y=0$.
Then the reality condition is  that the field is real $[Y(\tau,\theta)]^\dagger = Y(\tau,\theta)$, and therefore 
implies that $(J^i_n)^\dagger= J^i_{-n}$ and  $(\bar{J}^i_n)^\dagger= \bar{J}^i_{-n}$. This Lorentzian reality condition then doesn't imply that the Wick rotated field $Y(z,\bar{z})$ with $z = e^r e^{i\sigma}$ where $r=i\tau$, is a real field. Instead it implies that $[Y(z,\bar{z})]^\dagger = Y[\bar{z}^{-1},z^{-1}]$, which involves time reversal.
This CPT reality condition descends directly from having a Lorentzian field even if one works in an Euclidean framework.
In our case the field $X$ is a real solution of the Laplace equation, it doesn't descend from a Lorentzian equation, but satisfies from the onset an Euclidean equation of motion. The reality condition is $[X(z,\bar{z})]^\dagger = X(z,\bar z)$ instead. This is the correct reality condition for an Euclidean scalar field.

\section{Mode Expansion: The $k$ integer case}\la{Kint}
The goal is now to complete our analysis in the $k$ integer case. Holomorphicity implies that the scalar fields themselves decompose as a sum  $X^a(z,\bar{z})=X^a_+(z) +X^a_-(\bar z) $. 
As we are about to see, the structure of the zero modes depends drastically on wether the curvature is integer valued or not. Therefore for the reader's convenience we distinguish the two cases.
In this section we consider the case where $k\in \Z$, which includes the flat ($k=0$) case and which is in fact and quite remarkably equivalent to the case $k=0$. 
Then, we can introduce the mode expansion  
\ba\label{X}
X^a(z,\bar z)&=& x^a(z,\bar z) \\
&-&\sum_{n+k^a\neq 0}\frac{J^{a}_n z^{-n- k^a}}{(n+ k^a)} -
\sum_{n-k^a\neq 0} \frac{\bar{J}^{a}_n \bar{z}^{-n+k^a}}{(n- k^a)} \,,\n
\ea
where we have denoted the zero mode operator
\ba\la{x}
x^a(z,\bar z)&=& x^a +J^a_{-k^a} \ln z + \bar{J}^a_{k^a} \ln{\bar z}\n\\
&=&\tilde x^a+ \theta P^a
\,,
\ea
and we have defined the zero mode `position' and `momentum'
\ba
\tilde x^a&:=&x^a +(J^a_{-k^a}+\bar{J}^a_{k^a})\ln{r}\,,\\
P^a&:=&
i(J^a_{-k^a}-\bar J^a_{k^a})\,.\la{Pa}
\ea
From this expansion it is clear that the map $J\to\tilde{J}$, where 
$\tilde{J}_n^a = J_{n-k^a}^a$ and $\tilde{\bar{J}}^a_{n}= \bar{J}_{n+k^a}^a$, identifies the sector where $k$ is an integer with the sector where $k=0$. This is the reason behind the compactification of the connection space at the quantum level. 
\subsection{Symplectic structure}
We  assume that the puncture boundary $C_p$ is defined at constant radial coordinate $r$.
This assumption does not affect the generality of our construction, the reason is that, given any contour $C_p$, we can always choose the background metric $\eta$ entering the gauge fixing (\ref{gg}) so that $C_p$ is constant $r$. 
Let us also introduce the modes
\be\la{QJ}
Q^a_n= i(r^{-n-k^a}J^a_{n} - r^{n+k^a}\bar{J}^a_{-n})\,,
\ee
which represents the expansion of $e^a_\theta$ in terms of the $J^a_n, \bar{J}^a_n$, namely
\be\la{eexp}
e^a_\theta =  \sqrt{\frac{\kappa \gamma}{2\pi}} \left(\sum_n e^{-i\theta (n+k^a)}  Q^a_n\right)\,.
\ee
The unusual reality conditions on $J$  translates into a familiar one for these modes
\be
(Q_n^a)^\dagger =  Q^{\bar{a}}_{-n}.
\ee
Direct replacement of the expansion \eqref{X} in the symplectic form \eqref{symppp} shows that
\be \la{SymP}
\Omega=  \sum_a \left( \frac12 {\delta \chi^a \wedge \delta P^{\bar a}}  +i\sum_{n+k^a\neq 0} \frac{\delta {Q}^a_n\wedge \delta {Q}^{\bar a}_{-n} }{2(n+k^a)}\right)\,,
\ee
where we have defined the variable $\chi^a$ as
\ba 
\chi^a&:=&  \tilde x^a+ \sum_{n\neq 0 }\left( {J^a_{n-k^a}} +  {\bar{J}^a_{n+k^a}}
 \right) \frac{r^{-n}}{n}\,,
\ea
 which is conjugate to the momentum  \eqref{Pa}.
From the previous expression we can read the Poisson brackets 
\be\label{Qbra}
 \{Q_n^a,{Q}_m^b\} = -i{\delta^{a\bar{b}}}  (n+k^a) \delta_{n+m},\quad \{ \chi^a,P^{ b}\}= 2 {\delta^{a\bar{b}}}\,.
 \ee
We thus recover  the $U(1)^3$ Kac-Moody algebra \eqref{algebra2} plus a zero mode algebra. It is interesting to see that here the curvature appears as a twisting of the angular variables and that we also have, in general, a position variable $\chi^{\bar a} $ conjugated to $ P^a= Q_{-k^a}^a $. 

\subsection{Gauge and Dirichlet vs. Neumann boundary conditions }

Here we show that there is a relationship between the previous construction and the imposition of boundary conditions
at the punctures. Indeed no question about boundary conditions at $C_p$ arises if one deals with the symplectic structure in terms 
of the currents $Q^a_n$. However, if we decide to express the results of the previous subsection in terms of the currents $J^a_n$ and $\bar J^a_n$---more simply related via (\ref{J-X}) to the scalar fields $X^a$ on $\bar{D}$---then there is a direct link between boundary conditions for the $X^a$ and the parametrization of commutation relations.

First notice from \eqref{X} that the imposition of Dirichlet boundary conditions $\partial_\theta X^a=0$ at fixed $r$ (assumed to denote the value of the polar coordinate $r$ at a given puncture $C_p$) corresponds to the condition 
\be\la{Dir}
\delta Q_n^a=0\,.
\ee
Therefore, imposing Dirichlet boundary conditions results in killing all the degrees of freedom at the puncture.

This remark tells us that the symplectic structure just constructed admits degenerate directions. According to the standard theory, the latter must be considered as defining a gauge symmetry.
These gauge transformations can be written in terms of parameters $\alpha^a_n$ as follows
\be\label{residu}
\delta_\alpha J^a_n = r^{n+k^a} \alpha^a_n,\qquad \delta_\alpha \bar{J}^a_n = r^{{n-k^{ a} }} \alpha^a_{-n}.
\ee
These transformations preserve the reality condition as long as $(\alpha^{a}_{n})^\dagger = \alpha^{\bar a}_{-n}$.
Notice that this `Dirichlet gauge symmetry' can be written in terms of the frames as 
\be
z \delta e^a_z(z) =  \alpha^a(z/r),\qquad 
{\bar z} \delta e^a_{\bar z}(\bar z) =  \alpha^a(r/\bar{z})\,,
\ee
where $\alpha^a(x) = \sum_{n\in \Z} \alpha^a_n x^{-n}$.
It is immediate to identify a complete set of  gauge invariant observables under (\ref{residu}): they are given precisely by the currents
$Q^a_n$ defined in \eqref{Q}, i.e., the currents in the mode expansion of $e^a_\theta$.
Therefore, the diagonalization of $\Omega$ achieved in (\ref{SymP}) corresponds to expressing the 
symplectic structure in terms of the physical degrees of freedom $Q_n^a$  obtained by modding out the symmetry \eqref{residu}.

Another possibility to recover \eqref{Qbra} is to gauge fix the gauge symmetry (\ref{residu}) by imposing  a good gauge fixing condition.
It turns out that this can be naturally achieved by imposing Neumann boundary conditions $\pa_r X^a=0$ at $C_p$.  Indeed, Neumann boundary conditions on \eqref{X} imply that
\be\la{Neu}
\delta(J^a_n+\bar J^a_{-n}r^{2(n+k^a)})=0\,.
\ee
One can then show from (\ref{QJ}) that in this gauge we have 
\be
2i J^a_n= r^{n+k^a} Q^a_n\,,
\ee
from which \eqref{Qbra} follows again. Therefore, the physical degrees of freedom $Q^a_n$
can be conveniently identified with those encoded in the fields $X^a$ if Neumann boundary conditions are imposed at punctures.
 We can thus think of the loop strings at the punctures as Neumann strings.

\subsection{Loop Gravity fluxes and spin networks} \label{Sloop}

The main object of interest in loop gravity is the integrated flux
\be\la{Sigma}
\Sigma_D(\alpha)= \frac1{2\kappa \gamma} \int_D [e, e]^a \alpha_a, 
\ee
where $\alpha= \alpha^a\tau_a$ is a periodic\footnote{Because we are in the untwisted frame $e$ and not in the hatted frame $\hat{e}$.} Lie algebra valued element.
We have seen in (\ref{ggaauu}) that, when supplemented with the bulk term $-  \frac1{\kappa \gamma}\int_M\rd_A\alpha^i \wedge \Sigma_i$, \eqref{Sigma} yields the Gauss constraint $G_D(\alpha)$ and satisfies an SU$(2)$ algebra. In this section we want to investigate whether the flux itself,  without the addition of the bulk term, satisfies a non trivial algebra.
As we can see from the computation of the gauge algebra, this is possible only when the SU$(2)$ rotation labelled by $\alpha$ leaves the connection fixed, that is only when $\rd_A \alpha=0$.
This leaves two cases: Either the curvature  $K=k \tau^3$ is such that $k \in \mathbb{Z}$  is an integer, in which case the solutions of 
 $\rd_A \alpha=0$ are simply that $\alpha^a = a^a e^{-ik^a\theta}$, where $a^a$ are constants. Or $k$ is not an integer in which case the only solution satisfying the periodicity and covariant constancy condition is given by $\alpha^a = a^3 \delta^a_3$.
In the case where the curvature is an integer we get SU$(2)$ as a symmetry algebra, while in the case where the curvature is not integer-valued the symmetry is broken down to U$(1)$.

Notice that $\alpha$ such that  $\rd_A \alpha=0$ is exactly the choice that insures that the fluxes (\ref{Sigma}) can be written entirely in terms of the boundary  components $Q_n^a$ once we use \eqref{ephi22-bis}. This is not entirely obvious since the expression (\ref{Sigma}) is written as a bulk integral involving both $e_r$ and $e_\theta$. In other words, for the choices of $\alpha$ such that $\rd_A \alpha=0$ the flux operator is gauge invariant under the Dirichlet gauge symmetry. We are now going to explicitly show how the SU$(2)$ symmetry is recovered in the case $k\in \Z$.

When the curvature is integer valued, the zero mode sector consists of three  positions $\tilde x^a$ given in \eqref{x} and three momenta
$P^a$ given in \eqref{Pa}, while the oscillator modes are labelled by $Q_n^a$ for $n+k^a \neq 0$.  In the following we define $\tilde{Q}^a(\theta):= \sum_{n\neq 0} \frac{\tilde{Q}_{n}^a}{n} e^{-in \theta}$, where we denoted $\tilde{Q}_n^a =Q_{n-k^a}^a$   
If we assume that $\rd_A\alpha=0$ and for $k\in \mathbb{Z}$  we can evaluate these  flux generators on the kernel of the staticity constraint. 
We introduce the spin angular momenta 
\ba
M^a =\frac{1}{4\pi} \oint_{\pa D} [\tilde{Q},\rd \tilde{Q}]^a   \rd\theta\,,
\ea
and we can check that  the integrand is periodic.
The flux operator then simply reads  as a sum of orbital plus spin angular momenta (see \cite{Freidel:2013bfa} for a similar calculation)
\ba\label{sam}
\Sigma_D(\alpha) 
&=& {\kappa\gamma}\left(\tfrac12 [\tilde{x}, P]^a  + M^a \right) a_a, 
\ea
where $a_a$ was introduced above when parametrizong the solutions of $\rd_A\alpha=0$.
The generators $M^a$ are given as an infinite dimensional  generalisation of the Schwinger representation 
$M^a = -{\epsilon^a\!_{bc}}\sum_{n\neq 0 } : \frac{ \tilde{Q}^b_{n}\tilde{Q}^c_{-n}}{2n }:$. More explicitly,
\ba
&& M^3=-i \sum_{n\neq 0}  \frac{:\tilde{Q}^+_n\tilde{Q}^-_{-n}:}{ n} \,,\\ 
&&M^\pm=\mp i \sum_m  \frac{:\tilde{Q}^3_m\tilde{Q}^\pm_{-m}:}{m}\,,
\ea
 and  it can be shown that they satisfy the complex basis SU$(2)$ algebra 
 \be
 [M^3,M^\pm]=\pm i M^\pm\,,\quad[M^+,M^-]=iM^3\,.
 \ee
This, together with \eqref{sam}, establishes the link between the flux $\Sigma^a$  and the string angular momentum along $\pa D$.
Notice that the non-commutativity of the loop gravity fluxes at the boundary is consistent with the boundary constraint  \eqref{bbconst} which can therefore be implemented in the context of the LQG bulk quantization.
This shows how the original $SU(2)$ gauge symmetry of loop gravity is implicitly hidden in the $U(1)^3$ twisted    
Kac-Moody symmetry and finally is recovered upon 
the implementation of the boundary Gauss constraint.

\subsection{The energy-momentum tensor}

 Notice that \eqref{TD} can be expanded in terms of the components $(T_{zz}, T_{z \bar z}, T_{\bar z\bar z}) $ of a symmetric stress tensor as 
 \be 
 T_D(v) =\frac1{2\pi}\oint_{\pa D} \left[\rd z ( v^z T_{zz} + v^{\bar z} T_{\bar z z}) 
 +\rd \bar z ( v^z T_{z\bar z} + v^{\bar z} T_{\bar z \bar z})  \right],\n
 \ee
 where the components of the energy-momentum tensor are quadratic in the frame field. This corresponds to the usual expression for the Hamiltonian generator as 
 $
 H(v)=\int T_{\mu\nu} \xi^\mu \rd \Sigma^\nu,
 $ with components
 \be
 T_{zz} = \frac{\pi e_z^i e_{zi}}{\kappa \gamma} ,\qquad 
 T_{\bar z\bar z} = \frac{\pi e_{\bar z}^i e_{\bar zi}}{\kappa \gamma} ,\qquad
 T_{z\bar z} = \frac{\pi e_z^i e_{\bar zi}}{\kappa \gamma}. 
 \ee
 Since $g_{AB}= \eta_{ij} e_A^i e_B^j$ is the 2-dimensional metric of $S$, we see that the previous construction equates the 2-dimensional metric with the energy-momentum tensor: 
 \be 
 T_{AB} = \frac{\pi}{\kappa\gamma}  g_{AB}.
 \ee

\section{Mode Expansion: The $k$ non-integer case}\la{non-integer}

In the case where $k$ is not an integer, the mode expansion for the scalar fields reads
\ba
X^3(z,\bar z)&=& x^3(z,\bar z)-
 \sum_{n\neq 0 }\left( \frac{J^3_n}{nz^{n}} +  \frac{\bar{J}^3_n}{n\bar{z}^{n}}
 \right),
\\
 X^\pm(z,\bar z)&=&-\sum_{n\in \mathbb{Z}}\left( \frac{J^{\pm}_n}{(n\pm k)z^{n\pm k}} + \frac{\bar{J}^{\pm}_n}{(n\mp k)\bar{z}^{n\mp k}} 
\right)\,,
\ea
where the zero mode position is now only in the direction of the curvature:
\ba
x^3(z,\bar z)&=& x^3 +J^3_0 \ln z + \bar{J}^3_0\ln{\bar z}\n\\
&=&\tilde x^3+ \theta P^3
\,,
\ea
where we have defined
\ba\la{P3}
\tilde x^3&:=&x^3 +(J^3_{0}+\bar{J}^3_{0}) \ln{r}\,,\\
P^3&:=& i(J^3_{0}-\bar{J}^3_{0})=Q^3_0\,.
\ea
The symplectic form now reads
\be \la{SymP3}
\Omega=    \frac{\delta \chi^3 \wedge \delta P^{3}}{2}  +\sum_a\sum_{n+k^a\neq 0} i\frac{\delta {Q}^a_n\wedge \delta {Q}^{\bar a}_{-n} }{2(n+k^a)}\,,
\ee
where the effective position is
\ba 
\chi^3&:=&  \tilde x^3+ \sum_{n\neq 0 }\left( {J^3_{n}} +  {\bar{J}^3_{n}}
 \right) \frac{r^{-n}}{n}\,.
\ea
The SU$(2)$ symmetry is broken down to U$(1)$ in this case, with the U$(1)$ generator being simply by 
\be 
M^3=-i \sum_{n}  \frac{:Q^+_nQ^-_{-n}:}{ n+k} .
\ee

\section{Relationship with previous results}\la{SET}
In this Section we want to explain that some of the results we just established here were implicitly suggested
in previous work. 

\subsection{Quantum gravity at the corner}

Our present results shed light and clarity onto the previous work \cite{Freidel:2015gpa}. In fact now we see that in that paper one was actually scratching on the surface of the present results.  In that paper it was understood that the 
tangential frames $e_A^i=(e_z^i,e_{\bar z}^i)$ represent two types of data on $S$.
Using them one can reconstruct the metric $g_{AB}$ and the fluxes $\Sigma^i$, which are given by 
\be
g_{AB}= e_A^i e_{Bi},\qquad \Sigma^i=\frac12 \epsilon^{AB} [e_A,e_B]^i. 
\ee
The previous two sets of observables reproduce---from \eqref{cr}---an algebra that is isomorphic to $SL(2,\R)$ and
$SU(2)$ respectively. We now see that the $SL(2,\R)$ algebra is nothing else but the  (anomaly free) symmetry 
algebra constructed from the Virasoro generators $L_1,L_0$, and $L_{-1}$ in \eqref{Vir} at every punctures; while the $SU(2)$ algebra is the one 
explicitly recovered in \eqref{su2-1}. 
Moreover, we have shown that all these quantities have an interpretation in terms of a 3-dimensional string target fields $X^i$ (see for instance equations \eqref{L} and \eqref{schw}).
We have seen that the frames represents the holomorphic and anti-holomorphic string currents $(J^i,\bar{J}^i)$, and  that the tangential metric represents the string energy-momentum tensor $T_{AB}$. The full richness of the algebra found here  
was missed in that first investigation, but we appreciate now that the analysis was uncovering the right underlying structures.

\subsection{From Spin Networks to Virasoro states}
Let us first explain under the light of the present paper the nature of the results found in  \cite{Freidel:2009nu}  where a very natural question was investigated.  Given a spin network state $\Psi \in {\cal H}_{\bm{\vec\jmath}}$, where $\bm{\vec\jmath}=(j_1,\cdots j_N)$ and  ${\cal H}_{\bm{\vec\jmath}} = (V_{j_1}\otimes \cdots \otimes V_{j_N})^{\rm{SU}(2)}$ is the space of SU$(2)$ invariant vectors, with $V_j$ denoting the spin-$j$ representation, one can  express this vectir in the coherent state polarisation as an holomorphic functional 
\be
\Psi_{\bm{\vec\jmath}}(z_1,z_2,\cdots , z_{j_N}).
\ee
It was shown that the natural invariant scalar product on that space can be represented as an integral
\be
||\Psi_{\bm{\vec\jmath}}||^2 =\int \prod \rd^2 z_i \, K_{\bm{\vec\jmath}}(z_i;\bar{z}_i) \, |\Psi_{\bm{\vec\jmath}}(z_i)|^2 \,.
\ee
Now the main results of \cite{Freidel:2009nu} concern the properties of the Kernel $K_{\bm{\vec\jmath}}$. The remarkable fact proven there is that the integration Kernel $K_{\bm{\vec\jmath}}$ can be understood as the correlation function of an auxiliary CFT, namely
\be
 K_{\bm{\vec\jmath}} =\langle \varphi_{j_1}(z_1,\bar{z}_1)\cdots \varphi_{j_N}(z_N,\bar{z}_N) \rangle_{CFT}\,,
\ee
where $\varphi_j$ is a primary field of conformal dimension $\Delta_j =2(j+1)$ and of spin $0$.
Moreover, it was shown that this correlation function can be written as a Witten diagram \cite{Witten:1998qj}. In other words, it is given as the correlation function of a 3d  AdS-CFT:
\be
 K_{\bm{\vec\jmath}}(z_i,\bar{z}_i) = \int_{H_3}\rd^3 x \, G_{\Delta_j}(z_i,\bar{z}_i; x).
\ee
The integral here is  over the 3-dimensional hyperbolic space $H_3$. $G_\Delta(z,\bar{z}; x)$ is the bulk to boundary propagator of conformal weight $(\Delta,\Delta)$,  $x$ is  a point in the bulk and $(z,\bar{z})$ represent a point on the boundary of $H_3$.
$\Delta_j =2(j+1)$ is the conformal weight associated with the spin $j$.
What this formula expresses is the fact that changes of coherent state labels $z\to f(z)$ can be interpreted as a conformal transformation which is a symmetry of the spin network amplitude.
In other words, this result showed that the kernel of integration for spin networks carries a representation  the Virasoro algebra, a result which is now clear from the perspective developed here.

What we now understand with the results presented here  is that the label $z$  of the coherent states has a geometrical interpretation. It is not just a state label that one can chose arbitrarily. It represents a choice of the frame $e_z$ around the puncture of spin $j$ and it determines the shape of the metric. This means that we expect coherent state labels to now be acted upon by boundary operators like $e_z$ and $e_{\bar z}$ representing the tangential frame geometry.
If confirmed, this picture opens the way towards a new understanding of the dynamics of spin network states since one of the main roadblock in defining the Hamiltonian constraint was  the fundamental ambiguity in the determination of the frame field. This ambiguity is encoded, as we now understand, in the determination of the boundary operators, which is related to the choice of the coherent state label.

\subsection{Loops or no Loops ?}
{
Our analysis of boundary conditions has been deeply inspired by the approach developed in loop gravity in the sense that we have focused our analysis on background geometries that are concentrated around punctures.
However it should be clear by now that our analysis departs drastically from the traditional loop approach \cite{Ashtekar:2004eh}. In the traditional approach the flux operator is assumed to be  vanishing away from the puncture while here only the integrated flux---coming from \eqref{lala}---satisfies
\be
\Sigma_{\overline D} =\kappa\gamma \sum_p M_p^i \,,
\ee
where $M_p^i$ are the string angular momenta attached to each puncture.
This is so because in our case the flux density is determined by the simplicity constraint and reads $\Sigma^i =\tfrac12 [\rd X, \rd X]^i$, it {\it doesn't vanish} outside the punctures. The field which is taken to have a singular behavior and to vanish outside the punctures is  the curvature instead, which satisfies 
\be\label{curv2} F^i(A)(x) \overset{S}= 2\pi \sum_p K_p^i \delta^{(2)}(x,x_p).
\ee 
The fact that loop gravity admits another representation was first hinted by Bianchi in \cite{Bianchi:2009tj} (for an earlier consideration see \cite{simplest}). This point was established at the semi-classical level in \cite{Freidel:2011ue}, where it was  shown that  the phase-space of loop gravity  labels piecewise flat {\em continuum} geometry \cite{Freidel:2013bfa}  and that there exists two dual diffeomorphism invariant `vacuum' configurations: The loop vacuum $\Sigma=0$ or the spin foam vacuum $F=0$.
Then Dittrich et al. in \cite{Dittrich:2014wpa,Bahr:2015bra} proposed a quantization in which the dual vacuum $\hat{F}|0\rangle=0$ is implemented.
The success of loop gravity initially rests on the fact that the connection, which is the variable conjugated to $\Sigma$, is compactified at the quantum level in terms of holonomies, and that the vacuum state implementing $\hat\Sigma=0$ is therefore normalisable. This is not the case a priori for the dual vacuum: The variable $\Sigma$ is not compactified in a natural manner and the dual vacuum is not naturally normalisable. In the work by Dittrich et al. this fundamental problem was resolved either by resorting to a discretisation of space or in the continuum  by considering a discrete topology on the gauge group that induces a Bohr quantization and forces us to consider  only exponentiated fluxes.
 
 In our work we see that, on the one hand, the natural vacuum that follows from the study of gravity in the presence of boundaries  is indeed the one implementing $\hat{F}|0\rangle=0$ as postulated in those works.
 On the other hand, we  can infer  that this vacuum is a {\it normalisable} Fock vacuum carrying a representation of the Virasoro algebra.
 So the resolution of the Loop gravity conundrum of having a vacuum annihilating the curvature but still being normalisable is obtained here organically, without having to resort to the exotic Bohr compactification \cite{Ashtekar:2002vh}.
  The resolution lies in the presence of central charges, creating anomalies that allow the definition of normalisable Fock-like vacuum. These are compatible from the onset with a continuum formulation. This reconciles in spirit both  standard field  approaches with the loop gravity determination of describing gravity in terms of non-perturbative gauge-invariant observables.
The points sketched in this section deserve to be developed further, yet we consider this possible resolution an important aspect of our work.

\section{Discussion}\label{Sdis}

The results presented here are threefold: First we have identified a natural boundary term for first order gravity that generalises  the Gibbons-Hawking term {and} leads to a natural implementation of the simplicity constraint as a boundary equation of motion (see \cite{Wieland:2016exy} for a very recent implementation of the same idea in the  context of a null boundary).
Secondly,
we have shown that in the presence of a locally flat geometry there exists non trivial degrees of freedom  that can be attached to the punctures and whose origin is due to  the second class nature of the residual  tangent diffeomorphisms. We have also shown that these degrees of freedom carry a representation of a twisted $U(1)^3$ Kac-Moody symmetry, encoding a Virasoro algebra and an SU$(2)$ or broken U$(1)$ symmetry. These symmetries, attached at each puncture, generalise the SU$(2)$ algebra attached to each link in loop gravity into an infinite dimensional algebra with central charge $3$. Thirdly, we have shown that we can now represent at the quantum level and in terms of a Fock vacuum,  not only the flux operator $\Sigma_p$  but the triad $(e_z,e_{\bar z})$ itself. We have seen that the triad can be understood as the component of a current associated with 
string-like excitations living in a three dimensional internal  target space.

The possibility to represent the triad is  one of the most exciting outcome of this work, since it  may finally open up the possibility to define the Hamiltonian constraint of GR at the quantum level, at least on a subset of states, { in an ambiguity-free manner}.
Indeed, it is well known that the Hamiltonian constraint depends explicitly on the triad---not on the flux---and that within standard loop gravity, where the flux vanishes outside the punctures, the geometry of the triad  is totally ambiguous at best. This is the reason behind the huge quantization ambiguity that challenges the construction of an anomaly free dynamics in loop gravity \cite{Perez:2005fn, Dittrich:2012jq}. The infinite dimensional Virasoro representations attached to the tubular neighborhood of the punctures label the sets of possible frames around them and acts on the sets of admissible triads. It is now possible to think that we have enough control on the local degrees of freedom to construct more precisely the Hamiltonian constraint.

Another exciting opportunity that our work opens is the possibility to describe accurately the boundary states corresponding to a black hole in the semiclassical regime. Indeed, the CFT degrees of freedom discovered here could naturally account for the Bekenstein-Hawking area law in the context of LQG. The central feature that makes this possible in principle is the fact that the central charge of the CFT describing boundary degrees of freedom is proportional to the number of punctures that itself grows with the BH area. This is a feature that resembles in spirit previous descriptions \cite{Carlip:1998wz}. However, an important advantage of the present treatment is the precise identification of the underlying microscopic degrees of freedom.
A precise account of this  is work in progress and will be reported elsewhere \cite{in-progress}.

In our work the presence of Virasoro symmetries attached to punctures is related to the necessity of thickening the spin network graph into  tubular neighborhood (see Figure \ref{Figure}). 
Such thickening appeared first in the necessity of framing the Chern-Simons observables and  shown to be related to 
the presence of quantum group symmetries. It was postulated a long time ago \cite{Smolin:1995vq, Major:1995yz} that such a structure should appear in loop gravity and might be related to the presence of a non zero cosmological constant. This idea resurfaced recently and more precisely in the context of the computation of spin foam amplitudes in the presence of a background cosmological constant \cite{Haggard:2014xoa,Haggard:2015yda}. It would be interesting to relate these developments, as well as the recent emergence of quantum group structures in 2+1 LQG \cite{Noui:2011im, Pranzetti:2014xva, Cianfrani:2016ogm}, to the new framework presented here. 

Our results have another important implication. They tell us that there exists a very natural 
candidate for non geometric (matter) excitations that can naturally be coupled to the CFT degrees of freedom. The discovery of 
non-trivial degrees of freedom on boundaries strongly suggests that we could add extra degrees of freedom to the quantum geometry framework
not only at a boundary,  but also on the tip of open spin-networks in the bulk.
In the second scenario, these new degrees of freedom will be necessary to restore diff-invariance at the tip of open links in a way that is the analog 
of the standard Stueckelberg procedure \cite{Stueckelberg:1900zz}.  This is how spinning-particles (and fermion fields) are coupled to spin networks
in LQG. Our results suggest that, in addition to these familiar degrees of freedom, CFT excitations could live on the string-like defects defined by
open spin-network links.

Finally one of the most ramified aspect of our work is that 
it sheds new light on the nature of boundary degrees of freedom in the gravitational context.  First one sees that we have a precise realisation of the idea firmly established in  \cite{Donnelly:2016auv} that gauge theories in general and gravity in particular, possesses  physical boundary degrees of freedom that organise themselves under a representation of an infinite dimensional symmetry group.
We have identified here, in a particular context, the degrees of freedom as punctures and the algebra as containing a finite number of copies of the Virasoro algebra. 
From the general perspective of building a formulation of a generally covariant theory of finite regions \cite{Oeckl:2003vu}, our results show that CFT excitations around point-like defects are natural, and hence, expected to be an important part of the boundary 
data at the quantum level. These {\em boundary} degrees of freedom  are relevant when describing interactions (measurements) with finite
spacetime regions, as the gauge-dependent vector potential is essential in describing the coupling of electromagnetism with charged particles, which could 
represent detectors on a boundary or physical boundary conditions such as having a box made of conducting (free charges) plates \cite{Susskind:2015hpa}.
The possible advantage of this  perspective has been put forward some time ago \cite{Smolin:1995vq, Carlip:1998wz, Rovelli:2001bz, Rovelli:2013fga}. However, 
until recently it has been unclear how to encode boundary degrees of freedom on time-like or null surfaces in the quantum theory (especially when the quantum theory is defined in terms of Ashtekar-Barbero variables). Our work  shows, on the one hand, the importance of adding the appropriate boundary term to the action (\ref{uno}) which, in the case where the staticity constraint is satisfied, grants 
the conservation of the symplectic structure  evaluated on the space-like component of the boundary, where the usual construction of the LQG variables can be done.  
On the other hand, our work makes explicit the nature of the boundary degrees of freedom. These are two important features of the present work which appears as a step in the right direction in the definition of the quantum theory in open finite spacetime regions.

More broadly, it would be interesting to understand how the precise construction presented here of boundary degrees of freedom carrying representations of the Virasoro group relates to the appearance of conformal symmetries at the asymptotic boundary of AdS space \cite{Brown:1986nw}.
The appearance of Witten diagram entering the AdS3/CFT2 correspondence 
in the spin network evaluation is particularly striking in that respect.
It is also tempting to conjecture in the light of \cite{Donnelly:2016auv} that the degrees of freedom revealed here, under the assumption that the boundary curvature is localised around punctures, are related in general to a deeper understanding of the nature of soft modes in gravitational background \cite{Strominger:2014pwa}.
}

\section*{Acknowledgments} 

DP wishes to acknowledge the Templeton Foundation for
the supporting grant number 51876. AP acknowledges the OCEVU Labex (ANR-11-LABX-0060) and the A*MIDEX project (ANR-11-IDEX-0001-02) funded by the ``Investissements d'Avenir" French government program managed by the ANR. Research at Perimeter Institute for Theoretical Physics is supported in part by the Government of Canada through NSERC and by the Province of Ontario through MRI. LF would like to thank Lee Smolin and Djordje Minic for friendly support and critical inputs.


\end{document}